\documentclass[%
 aip,
 jmp,%
 amsmath,amssymb,
preprint,%
]{revtex4-1}

\usepackage{graphicx}
\usepackage{dcolumn}
\usepackage{bm}

\begin{document}
\title{The Role of Pressure in Inverse Design for Assembly}

\author{Beth A. Lindquist}
 \email{bethl@lanl.gov}
\author{Ryan B. Jadrich}
\affiliation{Theoretical Division, Los Alamos National Laboratory, Los Alamos, New Mexico 87545, USA}
\affiliation{McKetta Department of Chemical Engineering, University of Texas at Austin, Austin, Texas 78712, USA}
\author{Michael P. Howard}
\affiliation{McKetta Department of Chemical Engineering, University of Texas at Austin, Austin, Texas 78712, USA}
\author{Thomas M. Truskett} 
\affiliation{McKetta Department of Chemical Engineering, University of Texas at Austin, Austin, Texas 78712, USA}
\affiliation{Department of Physics, University of Texas at Austin, Austin, Texas 78712, USA}

\date{\today}

\begin{abstract}
Isotropic pairwise interactions that promote the self assembly of complex particle morphologies have been discovered by inverse design strategies derived from the molecular coarse-graining literature. While such approaches provide an avenue to reproduce structural correlations, thermodynamic quantities such as the pressure have typically not been considered in self-assembly applications. In this work, we demonstrate that relative entropy optimization can be used to discover potentials that self-assemble into targeted cluster morphologies with a prescribed pressure when the iterative simulations are performed in the isothermal-isobaric ensemble. By tuning the pressure in the optimization, we generate a family of simple pair potentials that all self-assemble the same structure. Selecting an appropriate simulation ensemble to control the thermodynamic properties of interest is a general design strategy that could also be used to discover interaction potentials that self-assemble structures having, for example, a specified chemical potential.
\end{abstract}

\maketitle

\section{\label{sec:intro}Introduction}

Inverse design strategies are powerful tools for discovering simple interactions that promote spontaneous assembly of particles into states exhibiting specific structural motifs.~\cite{IDSA1,IDSA2,IDSA3,IDSA4,IDSA5,IBI_clusters,IBI_RE_pores,IBI_pores2,RE_crystals,RE_clusters_pores_crystals,binaryRE,RE_FK,RE_crystals_kspace,RE_strings,IDpatch1,IDpatch2,IDpatch3,IDpatch3,IDpatch4,IDpatch5,IDtemplate,IDshapes,IDcubes,IDBCP,IDproperties,Ferguson_review,Jaeger_perspective,AlexanderKatzReview} Recently introduced approaches build upon molecular coarse-graining methods such as iterative Boltzmann inversion (IBI)~\cite{IBI_and_RE_ID,IBI2,IBI3} or relative entropy (RE) optimization,~\cite{IBI_and_RE_ID,RE_ID,general_RE} which prescribe how to map a detailed model onto one with reduced degrees of freedom, e.g., replacing a multibody potential with a pairwise interaction or a group of atoms with a single bead. In design for assembly, this allows one to discover simple pair potentials that drive particles to sample equilibrium configurations that closely match those exhibited by a model with considerably more complex interactions designed to favor a target structure. Such inverse strategies have enabled the design of isotropic pairwise interactions that self-assemble particles into a rich variety of phases including equilibrium cluster fluids,~\cite{IBI_clusters,RE_clusters_pores_crystals,IBI_RE_pores} porous mesophases,~\cite{IBI_RE_pores,IBI_pores2,RE_clusters_pores_crystals} colloidal crystals,~\cite{RE_crystals,RE_clusters_pores_crystals,binaryRE,RE_FK,RE_crystals_kspace} and fluids and gels comprising particle strings (``colloidomers'').~\cite{RE_strings}

The simplest and most na\"{i}ve coarse-graining strategy for determining such an interaction is direct Boltzmann inversion, which sets the optimized pair interaction between particles equal to the potential of mean force obtained from the target configurations.~\cite{IBI_and_RE_ID} Unsurprisingly, such optimized potentials are unable to adequately reproduce the structure of most targets of interest at equilibrium,~\cite{IBI_and_RE_ID} and more sophisticated, iterative coarse-graining strategies like IBI or RE optimization are required for successful design. Moreover, since the coarse-graining process changes the interparticle interactions, there is an inherent trade-off between reproducing the structural characteristics of the target and matching its corresponding thermodynamic properties.~\cite{IBIcorr_penalty,ibicorr1,ibicorr2,fmcorr1,fmcorr2,fmcorr3} For example, standard IBI seeks a pair potential that matches the radial distribution function $g(r)$ of the target, but not its pressure.~\cite{ibicorr2}
It is possible to largely preserve agreement between the structure of the optimized and target ensembles while also bringing their pressures into accord by incorporating a correction term into the interaction.~\cite{ibicorr1,ibicorr2,IBIcorr_penalty,fmcorr1,fmcorr2,fmcorr3} For instance, the pressure can be corrected in IBI by adding a linear ramp to the pair potential,~\cite{ibicorr1,ibicorr2} while force-matching (another coarse-graining strategy) has a different, volume-dependent pressure correction term.~\cite{fmcorr1,fmcorr2,fmcorr3} Often, the specific form of the optimized potential is not particularly important for molecular coarse-graining applications, and modifying the pair potential in these ways does not generally pose problems.

In the context of inverse design for self assembly, however, it may be necessary to constrain the functional form of the pair interaction in order to make contact with a particular experimental system or to learn about the fundamental requirements for a self-assembly process. The RE optimization framework is particularly well suited for this purpose because it optimizes the parameters of a specified potential. The form of the resulting potential, however, would not generally be preserved after application of a numerical pressure correction, motivating the need for other strategies to control the pressure for self assembly design applications. Fortunately, the RE framework is not specific to any particular statistical mechanical ensemble,~\cite{general_RE,RE_ID} and while RE optimization is often performed in the canonical ensemble, it can also be carried out in the isothermal-isobaric ensemble to yield an optimized interaction that possesses the desired pressure by construction. For colloidal self assembly, RE-optimized potentials are often thought of as effective interactions between colloids mediated by an implicit solvent; in this context, the corresponding osmotic pressure~\cite{osmotic} is controlled by the optimization. Perhaps more significantly, in applications of inverse design for assembly, the pressure can serve as a control parameter that can be tuned to bias the RE optimization toward a more net attractive (lower pressure) or repulsive (higher pressure) effective interaction, yielding a family of pair potentials that favor assembly of the same target morphology.

In this work, we demonstrate how the RE optimization strategy for inverse design is modified when simulated in the isothermal-isobaric ensemble. We validate the approach by revisiting a prior study where we optimized pairwise interparticle interactions in the canonical ensemble to assemble a rich variety of equilibrium mesophases (porous dispersions, lamallae, clusters, etc.).~\cite{IBI_RE_pores} For this article, we use the equilibrium cluster phases that emerged from this earlier study as our target ensemble. Clusters, characterized by self-limited growth and preferred finite aggregate size, are of fundamental interest in the context of self assembly~\cite{compint1,compint2,compint3,compint4,compint5,compint6,compint7,compint8,compint9,compint10,compint11,compint12,imptrep1,imptrep2,imptrep3,repcl1,repcl2,repcl3,repcl4,repcl5,repcl6,repcl7,repcl8,repcl9,CSD_1_and_imptrep,CSD_2_and_compint} but also are of practical relevance for understanding and controlling the viscosity of concentrated solutions of, e.g., therapeutic proteins for subcutaneous injection.~\cite{protein_clusters1,protein_clusters2,drug_delivery}

The balance of the article is organized as follows. In Sec.~\ref{sec:methods}, we give the RE update equation applicable when the coarse-grained simulations are performed in the isothermal-isobaric ensemble instead of the canonical ensemble, in addition to describing other relevant computational details. In Sec.~\ref{sec:results}, we first demonstrate that, for an appropriately chosen pressure, the protocol described in Sec.~\ref{sec:methods} recovers the same pair interaction as when the coarse-grained simulations are performed in the canonical ensemble. We then use the isothermal-isobaric optimizations to discover a family of cluster-forming potentials at different pressures and characterize the morphologies associated with the resulting interactions before concluding in Sec.~\ref{sec:conclusions}. 

\section{\label{sec:methods}Computational Methods}

\subsection{Relative Entropy Optimization in Other Ensembles}
\label{subsec:re}

RE optimization is equivalent to maximizing the likelihood that a given simulation protocol will sample the configurations characteristic of a target ensemble.~\cite{RE_ID,general_RE,max_likelihood,RE_crystals,RE_clusters_pores_crystals} The protocol is specified by both the total interaction potential $U($\textbf{R}$|\boldsymbol{\theta})$ (where $\boldsymbol{\theta}$ denotes a set of adjustable parameters and $\textbf{R}$ indicates a particle configuration) \emph{and} the thermodynamic ensemble in which the optimization simulations are carried out. When the target ensemble has a constant number of particles $N$ and volume $V$, the canonical (``NVT'') ensemble is a natural choice for the RE simulation protocol, but other ensembles are also permitted. In the present work, we perform the optimization simulations in the isothermal-isobaric (``NPT'') ensemble, where the pressure $P$ is constant, and the volume fluctuates. Our target ensemble is at constant volume, so its volume distribution is effectively a delta function. The RE optimized model will approach a similar volume distribution to avoid poor overlap of configurations between the two ensembles; the pressure will match the desired value by construction. 

We previously provided a derivation of the RE update scheme when the optimization simulations are performed in the canonical ensemble.~\cite{RE_crystals} The derivation for other ensembles proceeds analogously. The primary differences are 1) the probability of observing a given configuration $\textbf{R}$ depends on the ensemble, 2) the averages that are taken over the simulation data correspond to the chosen ensemble, and 3) prefactors that depend on $N$ or $V$ in the update scheme may need to be explicitly accounted for instead of absorbed into the optimization learning rate. When maximum likelihood fitting is performed in the NPT ensemble and the form of the potential is taken to be an isotropic pairwise interaction denoted $u(r|\boldsymbol{\theta})$, the parameters at step $i$ in the optimization, $\boldsymbol{\theta}^{(i)}$, are updated by

\begin{equation} \label{eqn:NPT_update}
\boldsymbol{\theta}^{(i+1)} = \boldsymbol{\theta}^{(i)} + \eta \Bigg[\int_{0}^{\infty} drr^{2}[\rho g(r|\boldsymbol{\theta},P)-\rho_{\text{tgt}} g_{\text{tgt}}(r|V_{\text{tgt}})]\boldsymbol{\nabla}_{\boldsymbol{\theta}}\beta u(r|\boldsymbol{\theta})\Bigg]_{\boldsymbol{\theta}=\boldsymbol{\theta}^{(i)}} ,
\end{equation}
where $\eta$ is the learning rate, $g$ and $g_{\text{tgt}}$ are the radial distribution functions of the trial system in the current step of the optimization at density $\rho$ and that of the target ensemble at density $\rho_{\text{tgt}}$, respectively, $V_{\text{tgt}}$ is the volume of the target ensemble, and $\beta=(k_bT)^{-1}$ where $k_b$ is Boltzmann's constant and $T$ is temperature.
A complete derivation for Eq.~\ref{eqn:NPT_update} can be found in the Appendix.

In prior work, we have either used a static value for the learning rate $\eta$ or infrequently adjusted $\eta$ manually in response to the behavior of the optimization. However, for optimizations in the NPT ensemble, we have found that automated adjustment of the learning rate is essential. The magnitude of the update to the potential is determined both by $\eta$ and by the difference between $\rho$ and $\rho_{\text{tgt}}$. As a result, a value of $\eta$ that might be appropriate when $\rho$ and $\rho_{\text{tgt}}$ are very close may yield an update that is so large that the optimization becomes unstable when $\rho$ deviates more strongly from $\rho_{\text{tgt}}$. Since $\rho$ can fluctuate frequently and rapidly over the course of an optimization, automated control over $\eta$ is needed to maintain stability. In the Appendix, we describe an empirical, automated procedure to update the learning rate in a way that leads to stable optimization. 

\subsection{Cluster Target Phase} 
\label{subsec:clusters}
To test the RE framework in the isothermal-isobaric ensemble, we optimized pair interactions to promote self assembly of clusters. By working at the relatively low densities where clusters readily form, we can use a large simulation box with a modest number of particles. Clusters have the additional benefit of being straightforward to characterize via the cluster size distribution (CSD).~\cite{CSD_1_and_imptrep,CSD_2_and_compint}

To compute the CSD, two particles were defined as neighbors if their interparticle separation was less than a specified cut-off distance, $r_{\text{cut}}$. Two particles are members of the same cluster if they are neighbors or if there is a contiguous pathway of neighboring pairs that connects them. The CSD is defined as the probability, $p(n)$, that an aggregate contained a specified total number of particles, $n$. The pairwise nature of this analysis can lead to artifacts, particularly for larger values of $r_{\rm cut}$. For instance, two aggregates that are largely spatially separated could be classified as a single cluster if a single particle bridges them. To address this, we resolved these groups of clusters by performing k-means clustering~\cite{kmeans1,kmeans2,kmeans3} on the unwrapped particle coordinates for the aggregate identified by the CSD calculation, where the number of individual clusters in a given group was taken to be the nearest integer to the quotient of the number of particles in the group by the most probable cluster size (primary peak in the CSD). The clustering was performed using the kmeans++ algorithm for selecting initial conditions.~\cite{kmeans++} 

From the identities of the clusters, we computed their centers of mass in order to determine the cluster-cluster radial distribution function, $g_{\text{cl-cl}}(r)$. We also determined which lattice (if any) was formed by the clusters using Polyhedral Template Matching (PTM) with a cut-off of 0.15 for the root-mean-square deviation relative to an ideal lattice.~\cite{ptm} PTM can distinguish between face-centred cubic (FCC), hexagonal close-packed (HCP), body-centred cubic (BCC), simple cubic, and icosahedral crystals and works well even at higher temperatures. The PTM calculations and visualization of the simulation configurations were performed using the Open Visualization Tool (OVITO), version 2.9.0.~\cite{ovito}

\begin{figure}[!htb]
  \includegraphics{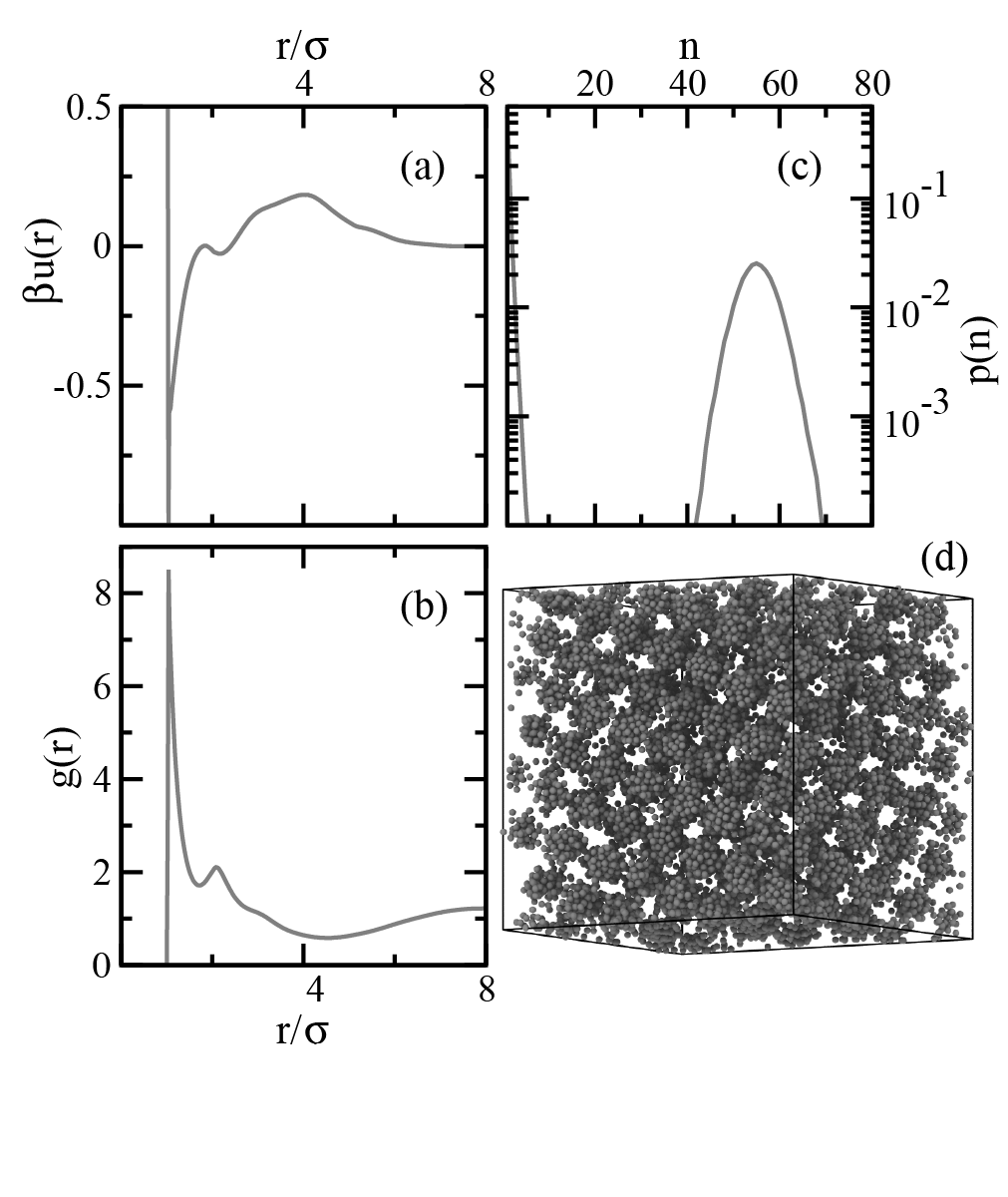}
  \caption{(a) The dimensionless isotropic pairwise interaction, $\beta u(r)$, discovered via inverse design using IBI in Ref.~\citenum{IBI_RE_pores}. This interaction drives self assembly of particles into various microphase-separated states as a function of number density ($\rho$) in the canonical ensemble. (b,c) The corresponding radial distribution function and cluster size distribution, respectively, at $\rho \sigma^3=0.144$. (d) A snapshot of clusters that assemble onto a lattice using the interaction shown in panel a at $\rho \sigma^3=0.144$. See Ref.~\citenum{IBI_RE_pores} for further simulation details.}
  \label{fgr:tgt_ensemble}
\end{figure}

The specific cluster target ensemble was inspired by prior work where we used IBI to discover an unconstrained pair interaction that prompted self assembly of a variety of microphase-segregated states as a function of $\rho$.~\cite{IBI_RE_pores} At $\rho \sigma^3$ values between 0.076 and 0.172 (where $\sigma$ is the nominal particle diameter), the IBI interaction (Fig.~\ref{fgr:tgt_ensemble}a) prompted self assembly of clusters. We generated the target ensemble by simulating this IBI-derived interaction at $\rho_{\text{tgt}} \sigma^3=0.144$ and computed $g_{\text{tgt}}(r)$ (Fig.~\ref{fgr:tgt_ensemble}b). The intra-cluster region extended out from $r=\sigma$ to approximately $r=3\sigma$, followed by a depletion zone between distinct clusters that led into cluster-cluster correlation peaks, the first of which occurred near $r=8\sigma$. The minimum at $r=1.71\sigma$ signaled the termination of the first coordination shell on the particle level, and therefore we used this value for $r_{\text{cut}}$ in the CSD calculations throughout. The target-ensemble clusters were relatively large and had good size-specificity (Fig.~\ref{fgr:tgt_ensemble}c); the median cluster size was 55 particles.  Additionally, the clusters themselves were arranged onto a lattice (Fig.~\ref{fgr:tgt_ensemble}d). PTM indicated that the structure had large co-existing domains of BCC and HCP lattices that account for 43\% and 40\% of the clusters, respectively.

\subsection{Functional Form for the Relative Entropy Optimization}
\label{subsec:fcn}

In our prior work,~\cite{RE_clusters_pores_crystals} the target ensembles were usually obtained from simulations with physically unrealistic or multi-body interactions, and RE optimization was used to replace these interactions with simpler pair potentials. Here, the target ensemble is obtained from an isotropic pair potential, but the form of the interaction is relatively complex and its features reflect many length scales. We can use RE optimization to replace this complex pair interaction with a more constrained functional form having simpler features.

To this end, we employed the following potential form in the RE optimization that was inspired by prior work~\cite{IBI_RE_pores,RE_clusters_pores_crystals} but has been further modified to reproduce some of the qualitative features of the IBI potential more closely. The potential had a hard-core-like contribution given by the Weeks--Chandler--Anderson (WCA) potential,
\begin{equation} \label{eqn:wca}
\beta u_{\text{\tiny{WCA}}}(r) = \left\{
\begin{array}{ll}
4 \bigg[ \bigg(\dfrac{\sigma}{r} \bigg)^{12} - \bigg( \dfrac{\sigma}{r} \bigg)^{6} \bigg] + 1, & r \leq 2^{1/6}\sigma \\ [15pt]
0, &  r > 2^{1/6}\sigma \\
\end{array} 
\right.,
\end{equation}
with $\sigma$ being the nominal particle diameter. The potential also had
an auxiliary interaction that furnished an attractive well and a repulsive barrier,
\begin{multline} \label{eqn:aux}
\beta u_{\text{\tiny{aux}}}(r) = \\ \left\{ 
\begin{array}{ll}
-\epsilon_{1} \bigg( \dfrac{-r + \sigma+\alpha_{1}+\alpha_{2}/2}{\alpha_1+\alpha_{2}/2} \bigg)^2 \textup{exp}\Bigg[-\bigg(\dfrac{r-\sigma-\alpha_{1}/2-\alpha_2/4}{\alpha_{1}/2+\alpha_2/4}\bigg)^8\Bigg] + \epsilon_2, & r \leq \sigma+a_1+a_2/2 \\ [15pt]
\epsilon_{2} \textup{exp}\Bigg[-\bigg(\dfrac{r-\sigma-\alpha_{1}-\alpha_{2}/2}{\alpha_{2}/2}\bigg)^2\Bigg], & r > \sigma+a_1+a_2/2
\end{array} 
\right.,
\end{multline}
where $\epsilon_1$ and $\epsilon_2$ are energy scales implicitly non-dimensionalized by the thermal energy and $\alpha_1$ and $\alpha_2$ are length scales. The first term provides an asymmetric attractive well that is deeper at smaller values of $r$ and smoothly connects to the peak of a Gaussian repulsive barrier given by the second term. The variables $\epsilon_1$, $\epsilon_2$, $\alpha_1$, and $\alpha_2$ are the adjustable parameters that comprise $\boldsymbol{\theta}$. 

To facilitate comparison between RE optimizations in different ensembles, we took special care to monitor the evolution of $\boldsymbol{\theta}$ to determine when the optimization was complete, as opposed to simply stopping the optimization when the desired structure was obtained. Some fluctuations in the parameters always persisted as a consequence of the stochastic nature of the simulations, but we manually identified the point at which the parameters no longer systematically evolved. We then selected the step from this region that had the best $g(r)$ matching, as quantified by the lowest value of $\gamma^{(i)}$ given by Eq.~\ref{eqn:conv_crit}.

\subsection{Simulation Details}
\label{subsec:sim_char_details}
All simulations were performed in \textsc{gromacs}, version 5.1.2.~\cite{GROMACS_1,GROMACS_2} The timestep was $\Delta t=0.001\sqrt{\beta m \sigma^2}$, where $m$ is the particle mass, and periodic boundary conditions were enforced in all three dimensions. A velocity-rescale thermostat was employed to control the temperature with a time constant of $\tau_T=100\Delta t$. The NPT simulations used a Berendsen barostat with a time constant of $\tau_{\rm P}=500 \Delta t$ and compressibility $7.2\times 10^{-4} \beta \sigma^3$ to maintain a constant pressure. The Berendsen barostat was chosen because it is particularly stable even when the starting box size is not consistent with the pressure of the barostat, which was often the case as the potential evolved over the course of the optimization. The disadvantage of this barostat is that it does not yield the exact NPT ensemble.~\cite{berendsen_issues} Therefore, we verified all NPT simulation results with NVT simulations after the optimization was complete, fixing the volume based on the average over the NPT simulation.

Since the simulations within the iterative optimization framework must be performed repeatedly (for each of the potentials presented here, hundreds of steps were generally required), we carried out relatively short and modestly sized simulations during the optimization. We then verified the results with larger and longer simulations outside of the optimization framework. During the optimization, the simulations contained 2,000 particles and were run for $10^6$ time steps each. The first $7.5\times10^5$ steps are taken to be the equilibration period, and any required quantities were measured over the remainder of the simulation. The NVT simulations were performed at $\rho_{\text{tgt}} \sigma^3=0.144$; the NPT simulations were also started from $\rho \sigma^3=0.144$ but $\rho$ evolved over the simulation. In order to further expedite the optimization simulations, we also used a somewhat aggressive value of $\approx 0.0025 k_b T/\sigma$ as the force threshold to determine the distance to truncate the pairwise interactions. 

In order to verify the results of the optimization, we characterized the optimized potentials using NVT simulations with 10,000 particles and $2\times10^7$ time steps, where the first $10^7$ time steps were defined as the equilibration period. Configurations were saved every $10^4$ steps for subsequent analysis. We also used a more conservative value of $0.001 k_b T/\sigma$ in the force to determine the truncation distance, effectively including more of the repulsive barrier in the simulation from Eq.~\eqref{eqn:aux}. As a result, the pressure measured from these NVT simulations was systematically slightly higher than the pressure input into the NPT optimization. To confirm that the pressure in the validation simulation was consistent with the NPT optimization, we computed a cutoff correction to the pressure,~\cite{allen} 
\begin{equation} \label{eqn:pressure_correction}
\Delta P = \frac{2 \pi}{3} \rho^{2} \int_{r_{\text{NPT}}}^{r_{\text{NVT}}} drr^{2} g_{\text{NVT}}(r|\boldsymbol{\theta}_{\text{opt}}) \Big( r \frac{du(r|\boldsymbol{\theta}_{\text{opt}})}{dr} \Big),
\end{equation}
where $\boldsymbol{\theta}_{\text{opt}}$ indicate the optimal parameters from the optimization, $r_{\text{NVT}}$ is the cut-off in the NVT validation simulation and $r_{\text{NPT}}$ is the cut-off from the NPT optimization. When the above term was included, the pressure from the validation simulation matched the targeted pressure to within 1.4\% in all cases, except for the lowest pressure optimization, which we will discuss in Sect.~\ref{sec:results}.

Finally, preliminary calculations indicated that using the final configuration from the previous step as the initial configuration in the iterative simulations caused kinetic issues with respect to the organization of the clusters. Clusters tended to become trapped in kinetically arrested and disordered configurations over the course of the optimization, whereas the clusters would facilely crystallize when initialized from a non-aggregated fluid state. Since the pressures associated with these two different arrangements of the clusters were different, the measured pressure from the NVT validation simulations did not match the pressure imposed by the barostat in the NPT simulations. In past work,~\cite{RE_crystals,RE_FK} we have used temperature annealing to avoid such issues, but large changes to the temperature at constant pressure change the box size significantly, which can cause practical issues in the simulation. Instead, we reset the initial configuration for every optimization simulation to a snapshot taken from an equilibrated WCA fluid simulation at $\rho_{\text{tgt}}$.

\section{Results and discussion}
\label{sec:results}

\begin{figure}[!htb]
  \includegraphics{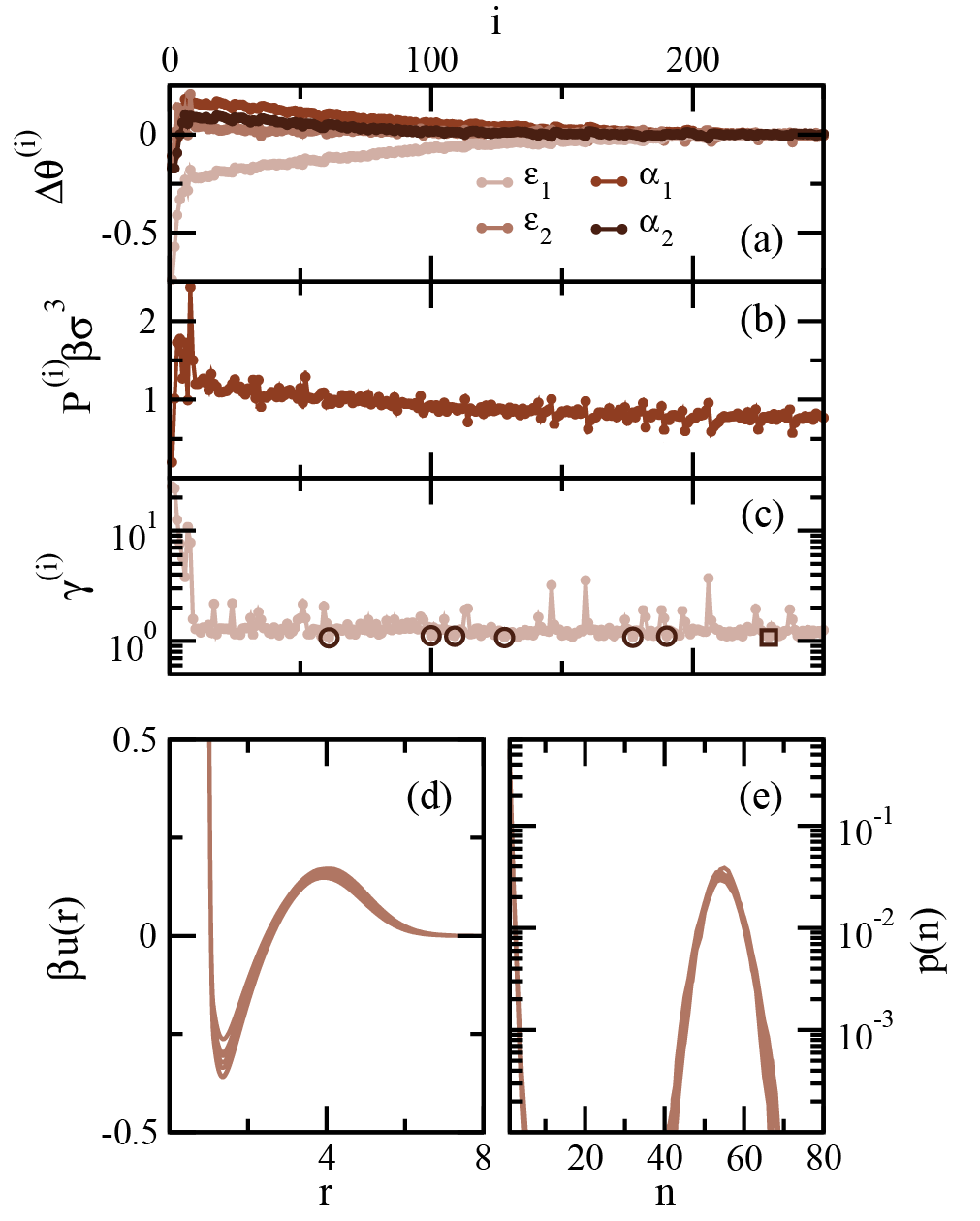}
  \caption{The evolution of (a) $\Delta \boldsymbol{\theta}^{(i)}=\boldsymbol{\theta}^{(i)}-\boldsymbol{\theta}_{\text{opt}}$, where $\boldsymbol{\theta}_{\text{opt}}$ are the optimal parameters, (b) the nondimensionalized pressure $P^{(i)}\beta \sigma^3 $, and (c) the convergence criterion $\gamma^{(i)}$ defined in Eq.~\ref{eqn:conv_crit} as a function of step $i$ in the NVT optimization at $\rho \sigma^3=0.144$. The step corresponding to the optimal interaction is denoted by the open square in panel (c). The (d) interaction potentials and (e) corresponding CSDs are shown for the self-assembled clusters for the steps denoted by the open symbols in (c).}
  \label{fgr:nvt_opt}
\end{figure}

To provide a basis for comparison for the NPT RE optimizations, we first performed a RE optimization in the NVT ensemble. Since the paths through parameter space may be different between the two ensembles, we monitored the convergence of the parameters to determine when the optimization was complete. Both the parameters (Fig.~\ref{fgr:nvt_opt}a) and the pressure (Fig.~\ref{fgr:nvt_opt}b) stopped evolving meaningfully after about 210 steps. We also show the convergence criterion at each step, $\gamma$, in Fig.~\ref{fgr:nvt_opt}c. Smaller values of $\gamma$ indicate better overlap between the target ensemble and the simulation performed with the optimized potential (see Eqn.~\ref{eqn:conv_crit}). We took the simulation having the lowest value of $\gamma$ as the final result of the optimization, denoted by the square in Fig.~\ref{fgr:nvt_opt}c.

Prior to the convergence of the parameters, we found that many values of $\gamma$ were comparable to that of the optimized potential, providing many interactions that appear consistent with the targeted cluster morphology. In Figs.~\ref{fgr:nvt_opt}d-e, we show a subset of these interactions and their corresponding CSDs from the steps demarcated by all the open symbols in Fig.~\ref{fgr:nvt_opt}c. The CSDs are all mutually consistent; therefore, we see that there is some flexibility in the interaction potential in terms of promoting self assembly of clusters with the correct size. However, the pressure associated with the potentials (Fig.~\ref{fgr:nvt_opt}b) varies from $P=0.76-1.01k_bT/\sigma^3$. 

\begin{figure}[!htb]
  \includegraphics{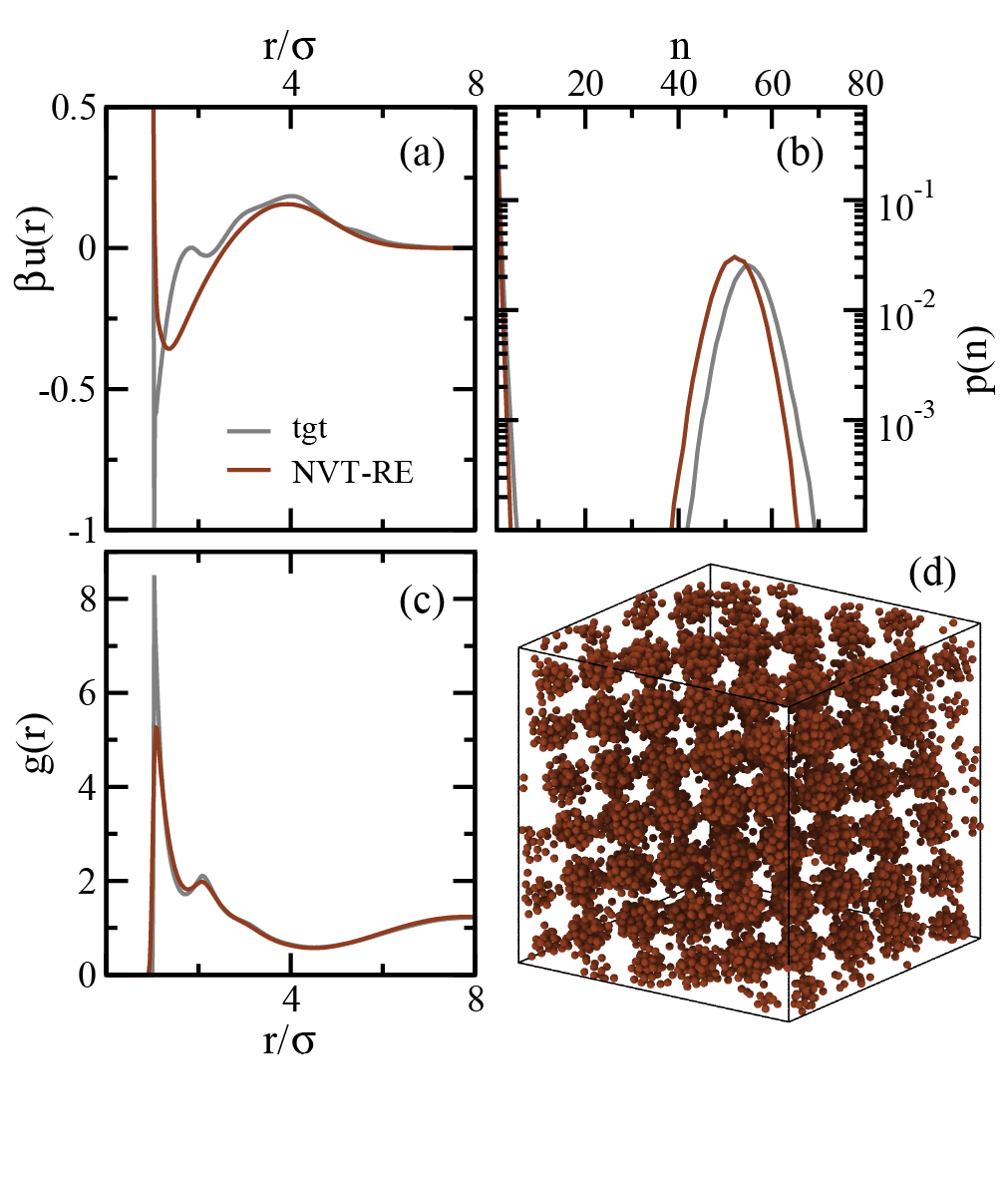}
  \caption{At $\rho \sigma^3=0.144$, comparison of (a) the IBI-derived interaction potential used to generate the target ensemble (gray) and the RE-optimized potential in the NVT ensemble with the constrained functional form given by Eq.~\ref{eqn:wca}--~\ref{eqn:aux} (dark orange), the (b) CSD and (c) $g(r)$ from the target ensemble and the optimized simulation. (d) A snapshot of clusters that are arranged onto a BCC lattice using the optimized RE interaction shown in panel a.}
  \label{fgr:nvt_v_tgt}
\end{figure}

The final optimized interaction and its corresponding CSD and $g(r)$ are shown in Figs.~\ref{fgr:nvt_v_tgt}a-c, where they are compared to their respective measurements from the target simulation. The restricted functional form of the interaction prevented perfect matching between the two, with the optimized interaction producing slightly smaller clusters than the target (the median values are at 52 and 55, respectively). This is reminiscent of prior work on porous mesophases, where a similarly restricted functional form produced slightly smaller pores than the unrestricted functional form.~\cite{IBI_RE_pores} The optimized $g(r)$ was slightly less structured overall, particularly near $r=\sigma$, which is probably a consequence of the deeper and sharper attractive well present in the target interaction. Such a sharp feature in the potential is not realizable for the constrained functional form. The optimized interaction prompted self assembly of clusters that crystallized onto a BCC lattice (Fig.~\ref{fgr:nvt_v_tgt}d). On average 70\% of clusters were classified as BCC. The coordination structure of most of the remaining clusters (25\% of all clusters) was unknown in the PTM analysis, which is likely at least partially attributed to defects in the lattice that occur due to the finite size of the simulation box. (There were slightly fewer than 200 clusters in the larger simulations.)

\begin{figure}[!htb]
  \includegraphics{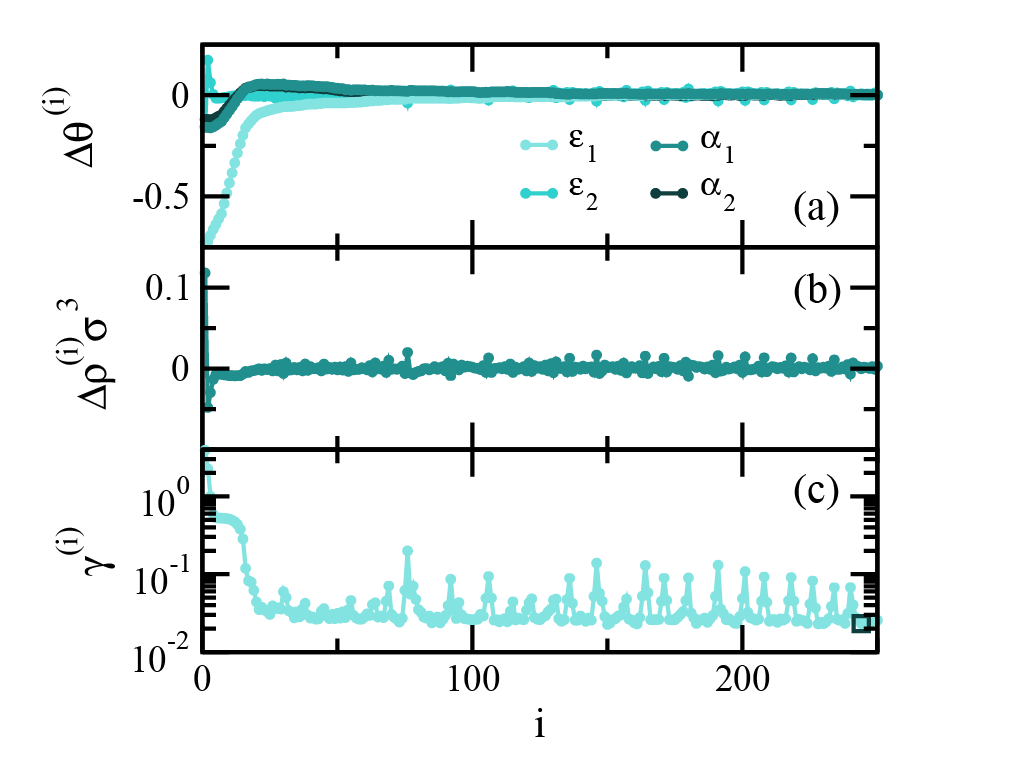}
  \caption{(a) The evolution of (a) $\Delta \boldsymbol{\theta}^{(i)}=\boldsymbol{\theta}^{(i)}-\boldsymbol{\theta}_{\text{opt}}$, where $\boldsymbol{\theta}_{\text{opt}}$ are the optimal parameters, (b) $\Delta \rho^{(i)} \sigma^{3} = \rho^{(i)} \sigma^{3}-\rho_{\text{tgt}} \sigma^3$, and (c) the convergence criterion $\gamma^{(i)}$ defined in Eq.~\ref{eqn:conv_crit} as a function of step $i$ in the NPT optimization, where $P = P^* = 0.788 k_b T/ \sigma^3$.}
  \label{fgr:npt_opt}
\end{figure}

In order to compare between RE in the NVT and NPT ensembles, we measured the pressure associated with the optimized interaction in Fig.~\ref{fgr:nvt_v_tgt}a. We extended the simulation with the optimized interaction for another $10^6$ timesteps, from which we computed a pressure of $P^*=0.788 k_b T/\sigma^3$. Using this value to control the pressure and setting $\rho_{\text{tgt}} \sigma^3$ equal to $0.144$, we carried out a RE optimization in the NPT ensemble, the results of which are shown in Fig.~\ref{fgr:npt_opt}a-c. While the parameters in Fig.~\ref{fgr:npt_opt}a stopped evolving more quickly in the NPT ensemble than the NVT ensemble as a function of optimization step, we note that the learning rate evolved with step differently in the two ensembles. In the case of the NPT calculation, the optimization seemed to heavily favor approximate matching of the density $\rho$ to $\rho_{\text{tgt}}$ (Fig.~\ref{fgr:npt_opt}b). The step associated with the final optimized interaction is shown by an open square in Fig.~\ref{fgr:npt_opt}c. 

\begin{figure}[!htb]
  \includegraphics{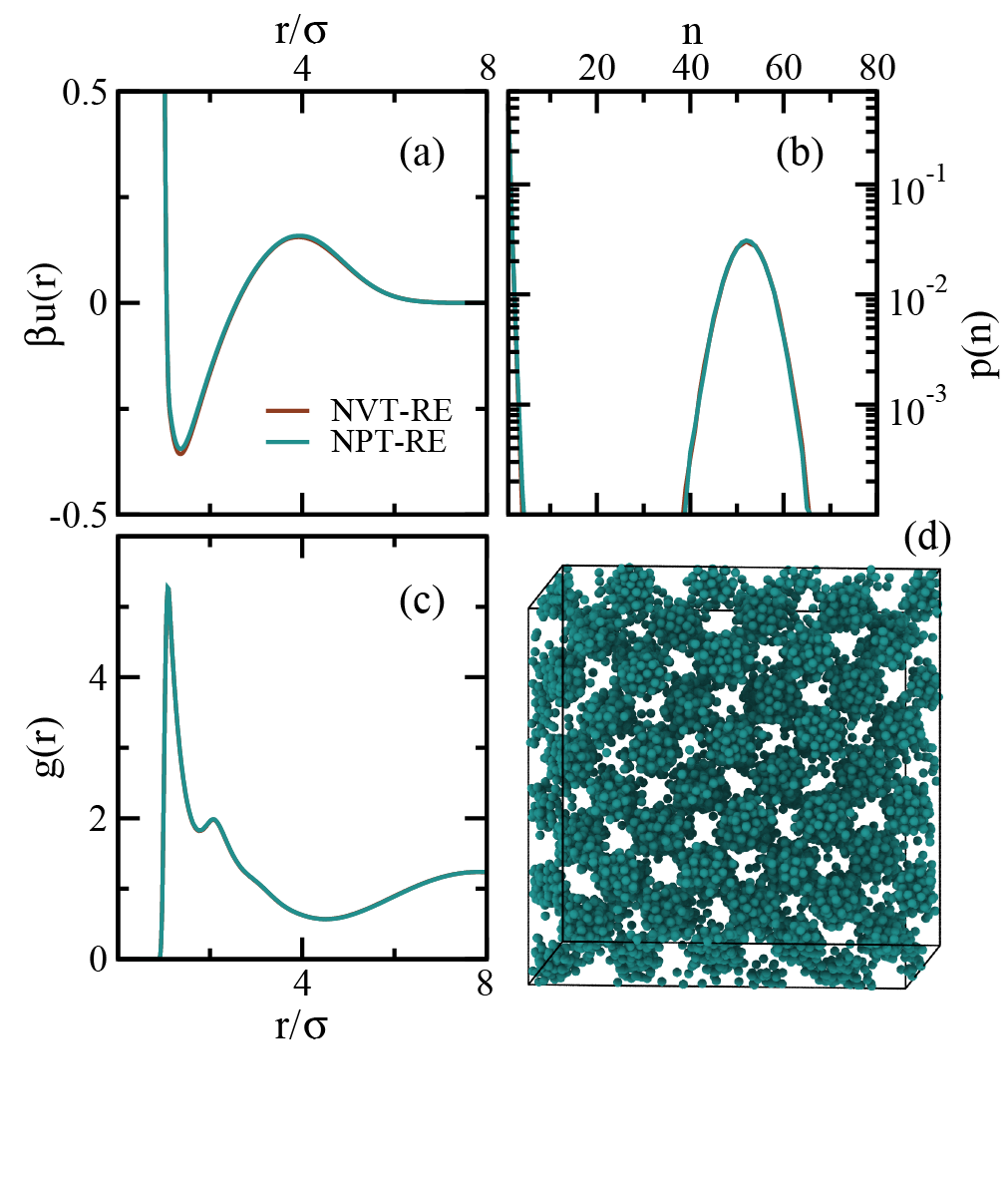}
  \caption{Comparison of (a) the optimized potential from the NVT ($\rho \sigma^3=0.144$, dark orange) and NPT ($P= P^* = 0.788 k_b T/ \sigma^3$, teal) optimizations, and the corresponding (b) CSD and (c) $g(r)$. (d) A snapshot of clusters that are arranged onto a BCC lattice using the optimized NPT interaction shown in panel a.}
  \label{fgr:npt_v_nvt}
\end{figure}

Despite the different paths through parameter space, the NVT and NPT RE optimizations arrived at effectively identical solutions. Each final optimized potential and the corresponding CSD and $g(r)$ are directly compared in Fig.~\ref{fgr:npt_v_nvt}a-c, where the results are nearly indistinguishable. The median cluster size was 52 particles in both cases. Both interactions promoted self assembly of the same morphology, with 72\% of the clusters from the NPT optimization assigned via PTM to a BCC structure (Fig.~\ref{fgr:npt_v_nvt}d) and 23\% of the clusters unclassified. As should be the case if the simulation box is sufficiently large, performing the optimization in the distinct ensembles but with compatible choices for the thermodynamic variables yielded equivalent results. 

\begin{figure}[!htb]
  \includegraphics{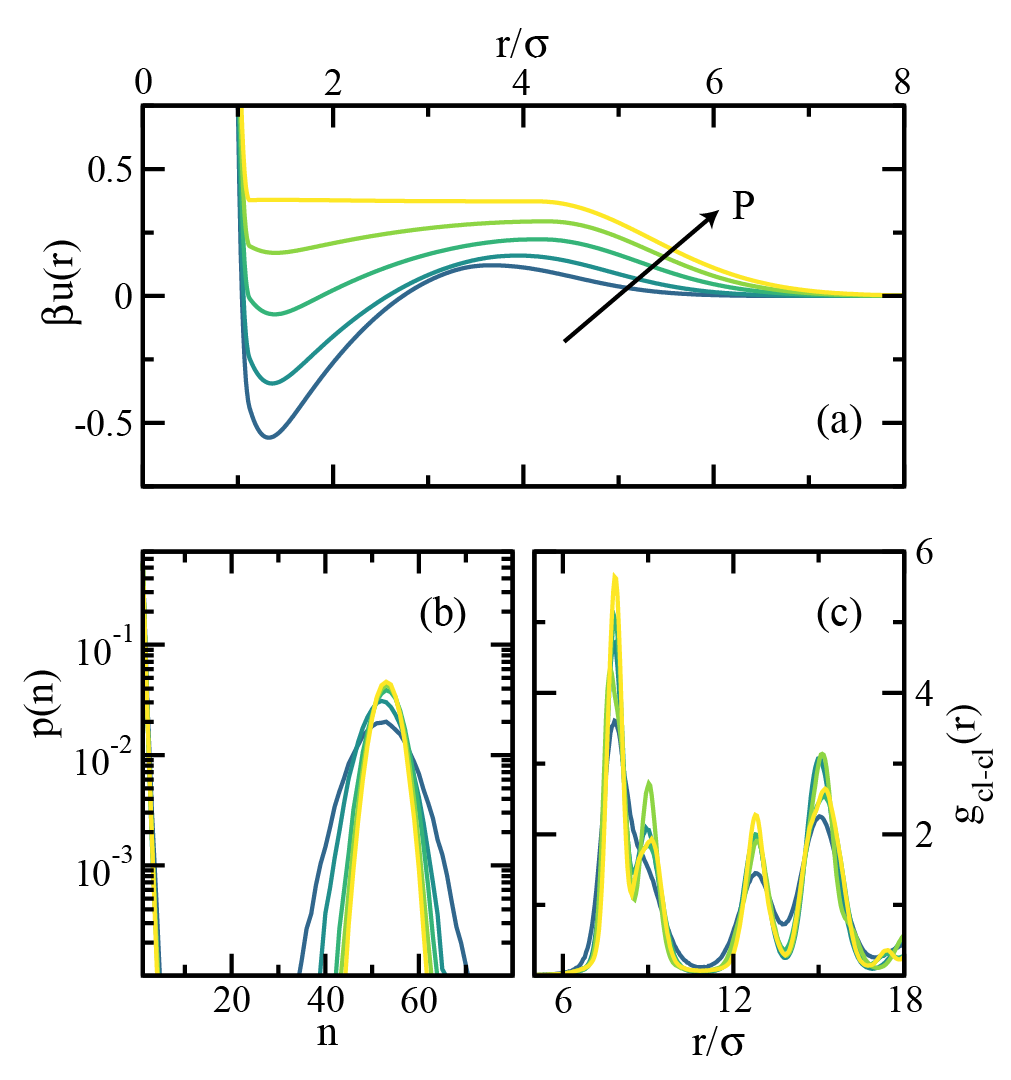}
  \caption{(a) The optimized interaction potentials, $\beta u(r)$, (b) the corresponding CSDs, and (c) the radial distribution functions between the center of masses of the clusters, $g_{\text{cl-cl}}(r)$, as a function of the pressure of the barostat in the NPT optimization for $P=0.5P^*$, $1.0P^*$, $2.0P^*$, $3.0P^*$, and $4.0P^*$, where the lighter colors indicate higher pressure and the arrow is drawn in the direction of increasing pressure.}
  \label{fgr:higherP}
\end{figure}

After validating the NPT optimization strategy, we subsequently varied the pressure to discover different interaction potentials that also resulted in a clustered microphase. The results for $P$ from $0.5P^*$ to $4.0P^*$ are shown in Fig.~\ref{fgr:higherP}a-c. As expected, potentials optimized at $P < P^*$ had a deeper attractive well and a more muted repulsive barrier; the inverse is true for $P > P^*$. On the whole, there is significant flexibility with respect to the parameters if either the volume or the pressure of the system can be tuned, though it is reasonable to assume that the parameters must be optimized in a coordinated fashion to produce the desired morphology. The CSDs resulting from the potentials in Fig.~\ref{fgr:higherP}a are shown in Fig.~\ref{fgr:higherP}b, where we see that the fidelity with respect to cluster size is consistent with the NVT optimization (median cluster size of 52 or 53 particles in all cases), with polydispersity decreasing as the pressure increases.  

All of the above interactions resulted in clusters that primarily crystallized onto a BCC lattice. PTM indicated that between 55-72\% of the cluster centers were in a BCC environment, and most of the remaining clusters were not classified as crystalline by PTM. The pair distribution functions of the cluster center-of-masses, $g_\text{cl-cl}(r)$, (Fig.~\ref{fgr:higherP}c) were also consistent with self assembly into a BCC lattice. 

From the $P=4.0P^*$ result, we note that attractions are not required to form a clustered phase, consistent with prior work.~\cite{repcl1,repcl2,repcl3,repcl4,repcl5,repcl6,repcl7,repcl8,repcl9} Analogously to previous work on pores and crystals comprised of single particles,~\cite{RE_clusters_pores_crystals,RE_FK} the attractive well can be completely replaced with a shoulder while retaining the targeted morphology. While cluster-forming potentials possessing competing interactions and those characterized by purely repulsive interactions are sometimes discussed as two distinct classes of potentials, by tuning the pressure, we see that there is a continuous family of interactions that spans this space. We did not perform calculations at a higher pressure than $P=4.0P^*$ because we found that the optimization did not converge (i.e., the parameters did not stop evolving) even after 1000 iterations. This may be due to some frustration in the optimization: the functional form qualitatively changes as $\epsilon_1$ becomes negative, i.e., the part of the function that generates the attractive well inverts to form another repulsive barrier and in doing so generates a rather abrupt attractive well between this barrier and the WCA component. Negative values for $\epsilon_1$ are disfavored in the optimization so that at sufficiently high pressures $\epsilon_1$ stays close to zero and other parameters must compensate to maintain the appropriate box size for the given pressure. Other functional forms tailored to generate purely repulsive potentials may be a better choice for higher pressures than those investigated here.

\begin{figure}[!htb]
  \includegraphics{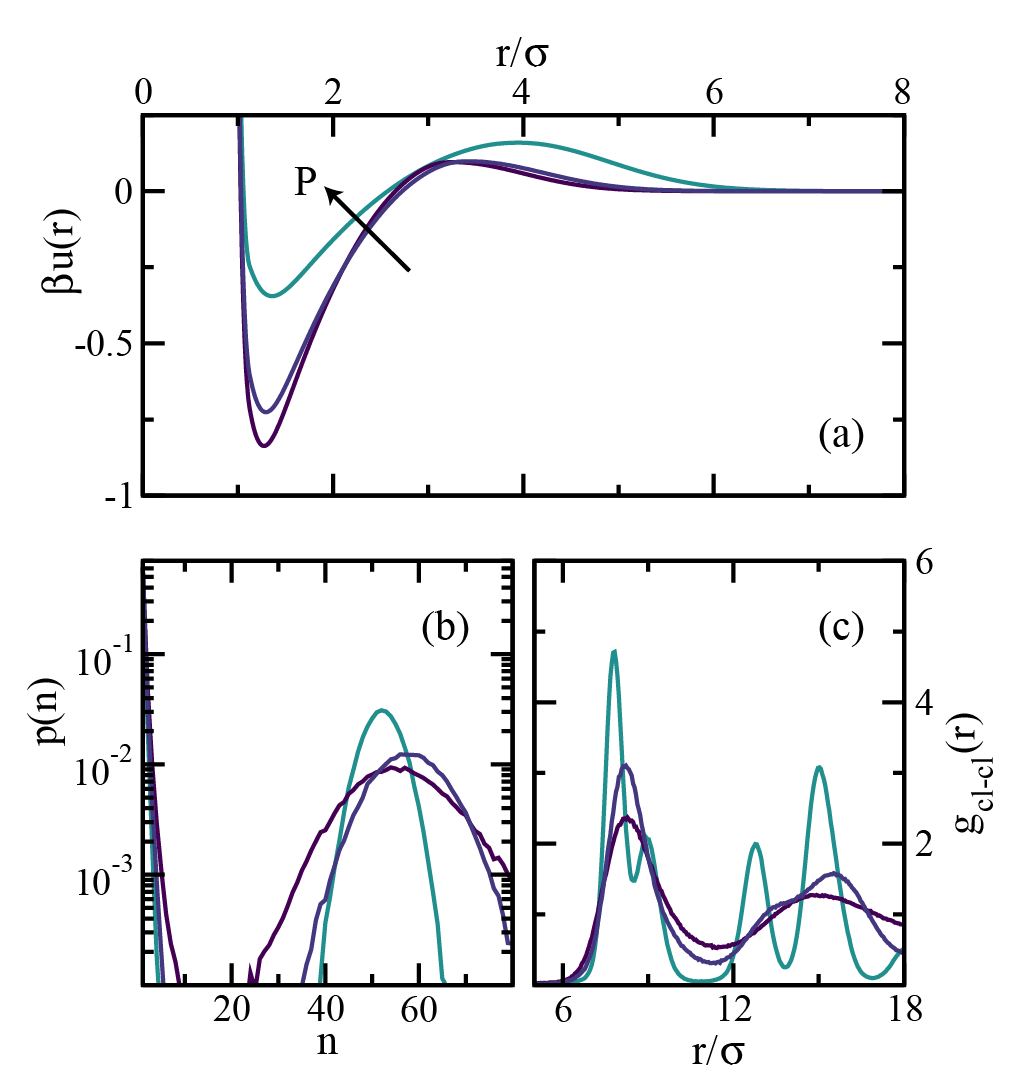}
  \caption{(a) The dimensionless optimized interaction potentials, $\beta u(r)$, (b) the corresponding CSDs, and (c) the radial distribution functions between the center of masses of the clusters, $g_{\text{cl-cl}}(r)$, as a function of the pressure of the barostat in the NPT optimization for $P=0.125P^*$ and $0.25P^*$, where $P=1.0P^*$ is shown for comparison. The lighter colors indication higher pressure and the arrow is drawn in the direction of increasing pressure.}
  \label{fgr:lowerP}
\end{figure}

We also performed NPT optimizations at even lower pressures than shown in Fig.~\ref{fgr:higherP}. The results for $P=0.125P^*$ and $0.25P^*$ are in Fig.~\ref{fgr:lowerP}, where the $P = 1.0P^*$ result is also plotted for context. We see a similar deepening of the attractive well as the pressure decreases in Fig.~\ref{fgr:lowerP}a. Conversely, while the repulsive barrier moves in to lower $r$ with lowering pressure, the magnitude of the repulsive barrier no longer significantly decreases from $P=0.25P^*$ to $P=0.125P^*$. It is understood that for interactions possessing competitive attractions and repulsions that a minimum degree of repulsion is needed to form clusters in order to thwart macroscopic phase separation,~\cite{imptrep1,imptrep2,imptrep3} and the form of the $P=0.125P^*$ potential appears to be influenced by this limitation. 

Perhaps as a consequence of the above considerations, the self-assembled structures corresponding to the lower pressure potentials in Fig.~\ref{fgr:lowerP}a had some notable qualitative deviations from the optimizations at higher pressures. First, the primary peak in the CSD was shifted to slightly larger clusters, and the clusters were significantly more polydisperse (Fig.~\ref{fgr:lowerP}b). For the case of $P=0.125P^*$, the clusters never crystallized during the validation simulation, continuing to exist as a fluid. (The structure of over 99\% of the clusters was not identifiable in the PTM analysis.) At $P=0.25P^*$, the assembly did order somewhat (though a plurality of the clusters, 46\%, were unclassified by PTM) into a polymorphic structure with significant degrees of BCC, HCP, and FCC crystal structures (34\%, 12\%, and 8\% of the clusters, respectively). The $g_\text{cl-cl}(r)$ for both low pressure optimizations are shown in Fig.~\ref{fgr:lowerP}c, where the morphological differences are apparent.

We also noticed that at the lower pressures the optimizations took noticeably longer to converge, with the $P=0.25P^*$ optimization requiring about 500 iterations (200--250 iterations was typical at the higher pressures) and the parameters in the $P=0.125P^*$ optimization failing to converge after 1000 iterations (though the convergence criterion did not improve over the last $\approx 800$ steps). Additionally, while the optimal $\rho$ value measured from the higher pressure results generally matched $\rho_{\text{tgt}}$ closely (less than 1\% difference), the final $\rho$ value measured from the optimal $P=0.125P^*$ simulation deviated more strongly (over 11\%). The discrepancy between the optimal $\rho$ value and $\rho_{\text{tgt}}$ posed a challenge to our optimization strategy since we began each simulation at $\rho_{\text{tgt}}$; this issue manifests as the pressure measured from the NVT validation simulation varying more strongly from the corresponding barostat pressure than in any other optimization ($\approx 7\%$ difference). By contrast, when we begin the NPT simulation from a disordered state at the correct density for the interaction, thereby allowing the clusters to self-assemble while $\rho$ is not evolving, we restore close agreement between the measured pressure of the NVT simulation and the barostat of the NPT simulation (within 1.5\%). Given that we observed kinetic issues when the iterative simulations were not initialized from a disordered state (see Sect.~\ref{sec:methods}), it seems reasonable that equilibration may also be inhibited if self assembly occurs while $\rho$ is also evolving, which could explain the above discrepancy with respect to pressure. 

On the whole, the low pressure regime investigated in this work underscores some of the limitations of tuning the pressure and still achieving self assembly of a targeted structure. Based on the above, we suggest that the convergence of the parameters and the closeness of $\rho$ for the optimized interaction to $\rho_{\text{tgt}}$ may serve as indicators for what pressures are compatible with a given targeted structure. On the other hand, there may be applications where tuning either the pressure or the interaction potential to modulate structural morphology may be desirable, and locating regions of phase-space where the morphology of the optimal self-assembled structures varies in response to pressure may be useful towards this end.  

The primary aim of this article has been validation of an approach to optimize for interactions that possess desired thermodynamic properties as well as structural correlations. In the course of doing so, we have also discovered a series of interaction potentials that self-assemble into clusters that may be useful in future work in the field of clustering. In Table 1 of the Appendix, we report the parameters that define all of the optimized potentials from this work and the corresponding $\rho$ values at which the validation simulations were performed.

\section{Conclusions}
\label{sec:conclusions}

\begin{figure}[!htb]
  \includegraphics{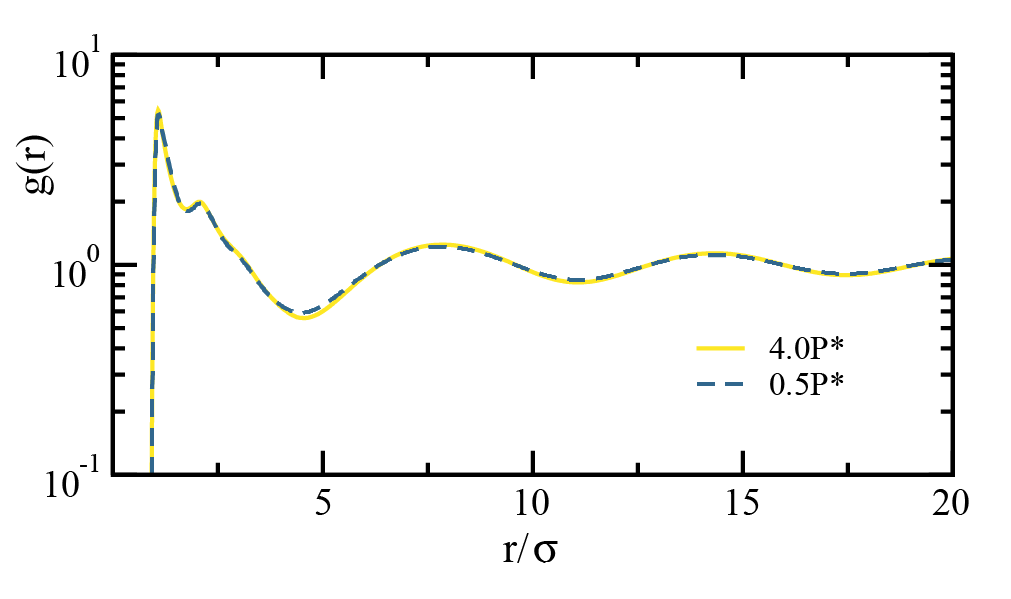}
  \caption{Comparison of the radial distribution functions $g(r)$ using the optimized interaction potentials from the $P=0.5P^*$ and $P=4.0P^*$ optimizations.}
  \label{fgr:gr_compare}
\end{figure}

In this article, we have demonstrated how, by manipulating the simulation protocol used in RE optimization, we can directly optimize for interaction potentials that possess a desired pressure while maintaining a constrained functional form. We showed that the RE optimization converges to the same solution in both the NVT and NPT ensembles when the pressure for the latter is chosen to be consistent with the output of the NVT optimization. Moreover, we have tuned the pressure in the NPT optimizations to generate a family of potentials that all form clusters of the correct size. As noted in prior work, specifying the pressure inherently degrades the matching between $g_{\text{tgt}}(r)$ and $g(r)$ when the potential is infinitely flexible;~\cite{ibicorr2} however, this is not necessarily the case when the functional form of the interaction is restricted. We found comparable $g(r)$ matching at all state points except for the two lowest pressure results that did not form a BCC lattice of clusters---these matched the target less well. In Fig.~\ref{fgr:gr_compare}, we compare the $g(r)$ computed from the $P=0.5P^*$ optimization to that from the $P=4.0P^*$ optimization. Despite the obvious differences in the optimized potentials (see Fig.~\ref{fgr:higherP}a), the radial distribution functions are well-matched, with the higher pressure result being only slightly more structured. 

Using a barostatted simulation protocol within the RE framework provides a straightforward and computationally convenient means to control the pressure associated with the interaction that is outputted by the RE optimization. This protocol allows us to discover the interaction form that is most likely to reproduce the target ensemble subject to the pressure constraint as opposed to tuning an interaction potential that matches the pressure in an ad hoc fashion. Finally, control of thermodynamic quantities via RE optimization is not restricted to the pressure. For instance, the chemical potential of a coarse-grained interaction could be tuned via performing the iterative simulations in the grand canonical ensemble.

\section{Acknowledgments}
This research was primarily supported by the National Science Foundation through the Center for Dynamics and Control of Materials: an NSF MRSEC under Cooperative Agreement No. DMR-1720595. This work was also partially supported by the Welch Foundation (F-1696). We acknowledge the Texas Advanced Computing Center (TACC) at The University of Texas at Austin for providing HPC resources. B.A.L. acknowledges support from the Darleane Christian Hoffman Distinguished Postdoctoral Fellowship at Los Alamos National Laboratory.

\setcounter{figure}{0}
\setcounter{equation}{0}
\renewcommand\thefigure{A\arabic{figure}}
\renewcommand{\thesection}{\thepart .\arabic{section}}
\renewcommand\theequation{A\arabic{equation}}

\renewcommand{\thesubsection}{\arabic{subsection}}

\section*{Appendix}
\label{sec:appendix}

\subsection{Relative Entropy Update Equation in the NPT ensemble}

The update scheme for RE optimization within the NPT ensemble follows analogous mathematical steps to the procedure outlined in prior work for the NVT ensemble.~\cite{RE_crystals} As before, we assume a target ensemble of $M$ statistically independent and identically distributed (IID) configurations, $\textbf{R}_{1:M}$, all of which have fixed volume $V_{\text{tgt}}$. The target ensemble should reflect the desired structural correlations to be realized via the optimized interaction. Adopting the NPT ensemble as the simulation protocol, the probability of observing a configuration ($\textbf{R}$) with a volume $V$ is given by the following Boltzmann factor
\begin{equation} \label{eqn:npt_boltzmann}
p(\textbf{R},V|\boldsymbol{\theta},P) \equiv \dfrac{\exp[-\beta U(\textbf{R}|\boldsymbol{\theta})-\beta PV]}{Z(\boldsymbol{\theta}, P)}
\end{equation} 
where $U(\textbf{R}|\boldsymbol{\theta})$ is the potential energy for configuration $\textbf{R}$, $\boldsymbol{\theta}$ is a vector of the tunable values that parametrize the potential, $\beta$ is the inverse thermal energy, $P$ is the pressure, and $Z(\boldsymbol{\theta}, P)$ is the partition function. Similarly, the  probability of observing the set of IID target configurations, $\textbf{R}_{1:M}$, with volume $V_{\text{tgt}}$, is given by the product of probabilities with the form of Eqn.~\ref{eqn:npt_boltzmann}:
\begin{equation} \label{eqn:likelihood}
p(\textbf{R}_{1:M},V_{\text{tgt}}|\boldsymbol{\theta},P) \equiv \prod_{i=1}^{M} p(\textbf{R}_{i},V_{\text{tgt}}|\boldsymbol{\theta},P) = \prod_{i=1}^{M} \dfrac{\exp[-\beta U(\textbf{R}_{i}|\boldsymbol{\theta})- \beta PV_{\text{tgt}}]}{Z(\boldsymbol{\theta},P)}
\end{equation} 
This quantity measures how likely a simulation in the NPT ensemble with parameters $\boldsymbol{\theta}$ is to sample the configurations $\textbf{R}_{1:M}$ at the target volume. Maximization of this quantity is called maximum likelihood fitting.~\cite{max_likelihood} 
	
As in the NVT derivation, it is easier to maximize the log-likelihood. Taking the natural log of Eqn.~\ref{eqn:likelihood} and dividing by $M$ yields:
\begin{equation} \label{eqn:log_likelihood}
\dfrac{1}{M}\text{ln}p(\textbf{R}_{1:M},V_{\text{tgt}}|\boldsymbol{\theta},P) = - \dfrac{1}{M}\sum_{i=1}^{M}\Big[\beta U(\textbf{R}_{i}|\boldsymbol{\theta})+ \beta PV_{\text{tgt}}\Big] - \text{ln}Z(\boldsymbol{\theta},P)
\end{equation}
which can be written as 
\begin{equation} \label{eqn:log_likelihood_2}
\langle\text{ln}p(\textbf{R},V_{\text{tgt}}|\boldsymbol{\theta},P)\rangle_{p_{\text{tgt}}(\textbf{R}|V_{\text{tgt}})} = -\langle\beta U(\textbf{R}|\boldsymbol{\theta})+ \beta PV_{\text{tgt}}\rangle_{p_{\text{tgt}}(\textbf{R}|V_{\text{tgt}})} - \text{ln}Z(\boldsymbol{\theta},P)
\end{equation}
in the large configuration limit (i.e., $M \rightarrow \infty$) where $p_{\text{tgt}}(\textbf{R}|V_{\text{tgt}})$ is the probability distribution of the target simulation. This expression only differs explicitly from the NVT version through the additional pressure-volume term.
	
Within the gradient ascent optimization algorithm, updates to $\boldsymbol{\theta}$ follow from
\begin{equation} \label{eqn:grad_descent}
\boldsymbol{\theta}^{(i+1)} = \boldsymbol{\theta}^{(i)} + \eta \big[\boldsymbol{\nabla}_{\boldsymbol{\theta}} \langle\text{ln}p(\textbf{R}, V_{\text{tgt}}|\boldsymbol{\theta},P)\rangle_{p_{\text{tgt}}(\textbf{R}|V_{\text{tgt}})}\big]_{\boldsymbol{\theta}=\boldsymbol{\theta}^{(i)}}
\end{equation}
where $\eta$ is the learning rate that is empirically tuned to maintain a stable optimization. Application of the gradient operator to Eqn.~\ref{eqn:log_likelihood_2} yields	
\begin{equation} \label{eqn:grad_log_likelihood}
\boldsymbol{\nabla}_{\boldsymbol{\theta}} \langle\text{ln}p(\textbf{R},V_{\text{tgt}}|\boldsymbol{\theta}, P)\rangle_{p_{\text{tgt}}(\textbf{R}|V_{\text{tgt}})} = -\langle \boldsymbol{\nabla}_{\boldsymbol{\theta}} \beta U(\textbf{R}|\boldsymbol{\theta})\rangle_{p_{\text{tgt}}(\textbf{R}|V_{\text{tgt}})} + \langle \boldsymbol{\nabla}_{\boldsymbol{\theta}} \beta U(\textbf{R}|\boldsymbol{\theta})\rangle_{p(\textbf{R},V|\boldsymbol{\theta},P)} 	
\end{equation}
which is formally equivalent to the NVT result, as the pressure-volume contribution in Eqn.~\ref{eqn:log_likelihood_2} is a constant with respect to $\boldsymbol{\theta}$. 

Differences due to the RE-optimization ensemble are only implicitly contained within the average that is performed over the simulation data (the right-most term in Eq.~\ref{eqn:grad_log_likelihood}). As implied by Ref.~\citenum{RE_ID}, the general relative entropy update in any other ensemble is given by the following replacements: $p_{\text{tgt}}(\textbf{R}|V_{\text{tgt}})\rightarrow p_{\text{tgt}}(\boldsymbol{R}|\boldsymbol{E}_{\text{tgt}})$ and $p(\textbf{R},V|\boldsymbol{\theta},P)\rightarrow p(\boldsymbol{R},\boldsymbol{E}|\boldsymbol{\theta},\boldsymbol{I})$ where $\boldsymbol{E}$ are the fluctuating extensive variables coupled to the fixed intensive variables $\boldsymbol{I}$, and $\boldsymbol{E}_{\text{tgt}}$ are the fixed values of the extensive variables in the target ensemble.
	
Eqn.~\ref{eqn:grad_log_likelihood} can be greatly simplified in the case of isotropic pair interactions. For a one-component system with particles interacting via the pair potential $u(r|\boldsymbol{\theta})$, the total potential energy is $U(\textbf{R}|\boldsymbol{\theta})\equiv \dfrac{1}{2}\sum_{i \neq j}^{N}u(r_{i,j}|\boldsymbol{\theta})$. Substituting this into Eqn.~\ref{eqn:log_likelihood_2}, and integrating over all of the coordinates except those in the pair potential yields
\begin{equation} \label{eqn:grad_log_likelihood_2}
\boldsymbol{\nabla}_{\boldsymbol{\theta}} \langle\text{ln}p(\textbf{R},V_{\text{tgt}}|\boldsymbol{\theta}, P)\rangle_{p_{\text{tgt}}(\textbf{R}|V_{\text{tgt}})} \\
= 2\pi N   \int_{0}^{\infty} drr^{2}[\rho g(r|\boldsymbol{\theta},P)-\rho_{\text{tgt}} g_{\text{tgt}}(r|V_{\text{tgt}})]\boldsymbol{\nabla}_{\boldsymbol{\theta}}\beta u(r|\boldsymbol{\theta})
\end{equation}
where $\rho$ and $\rho_{tgt}$ are the ensemble-averaged optimized and target densities, respectively, and $g(r|\boldsymbol{\theta},P)$ and  $g_{\text{tgt}}(r|V_{\text{tgt}})$ and are the corresponding radial distribution functions. Importantly, in writing Eqn.~\ref{eqn:grad_log_likelihood_2}, we have assumed that the pair interaction is finite in range and the system is macroscopic in size. This allows us to extend the integrals--which are technically over different volumes--arising from both terms in Eqn.~\ref{eqn:grad_log_likelihood} to infinity and add them together. This expression is similar to the NVT analogue apart from the different values of density attached to each radial distribution function. The factors in front of the integral are constants and can be absorbed into the learning rate of Eqn.~\ref{eqn:grad_descent}.

\subsection{Additional Details for Relative Entropy Optimization}

We use the evolution of the following convergence criterion 
\begin{equation} \label{eqn:conv_crit}
\gamma^{(i)} = \int_{0}^{\infty} drr^{2}[\rho g(r|\boldsymbol{\theta},P)-\rho_{\text{tgt}} g_{\text{tgt}}(r|V_{\text{tgt}})]^2
\end{equation}
which is an integrated measure of the similarity of the current step $i$ to the target ensemble, to govern the learning rate $\eta$. (See Eq.~\ref{eqn:grad_log_likelihood_2} for definitions of the symbols.) In particular, we compute $\delta^{(i)}=(\gamma^{(i-1)}-\gamma^{(i)})/\gamma^{(i-1)}$. If $\gamma$ significantly decreases from one step to the next (i.e. $\delta^{(i)} < -0.1$), then the optimization is converging as desired. Therefore, $\eta$ remains the same ($\eta^{(i+1)} = \eta^{(i)}$). If the convergence criterion increases, we have empirically observed that the optimization might be entering an unstable region in parameter space and $\rho$ may begin to fluctuate. If the increase in $\gamma$ is modest (i.e., $0.1 < \delta^{(i)} < 0.25$), then we maintain the same value for $\eta$ as above. However if $\delta^{(i)}$ is larger, then the learning rate is decreased ($\eta^{(i+1)} = 0.5 \eta^{(i)}$ if $0.25 \le \delta^{(i)} < 2.0$ and $\eta^{(i+1)} = 0.25 \eta^{(i)}$ if $\delta^{(i)} \ge 2.0$) to stabilize the optimization. Finally, if the $\gamma$ changes little ($-0.1 \le \delta^{(i)} \ge 0.1$), then it is possible that the optimization is not exploring parameter space efficiently, and so $\eta$ is increased accordingly ($\eta^{(i+1)} = 2 \eta^{(i)}$). We also include minimum and maximum values for $\eta$ ($0.001 \le \eta \le 0.16$ for most cases in this work), and we initialized the optimization using the minimum value for $\eta$.

For the initial guess for the NVT optimization and the subsequent NPT calculation where the pressure is set to the measured pressure from the preceding NVT result ($P^*$), we use a WCA potential (practically done by setting $\epsilon_1$ and $\epsilon_2$ equal to zero in Eq.~\ref{eqn:aux}). As can be seen from Eq.~\ref{eqn:aux}, the length scales $\alpha_1$ and $\alpha_2$ are non-linear parameters and therefore require reasonable initial guesses for the optimization to converge. Though the exact functional form differs, $\alpha_1$ and $\alpha_2$ have the same qualitative interpretation as in related prior work. Therefore, we use initial guesses for the length scales from prior RE calculations ($\alpha_1=1.46$ and $\alpha_2=2.55$).~\cite{IBI_RE_pores} Preliminary calculations demonstrated that the optimization did not appear sensitive to the choice of initial guess so long as the $\alpha$ values were reasonable. One other choice of initial guess was arrived at by fitting the functional form of Eqs.~\ref{eqn:wca}--~\ref{eqn:aux} to the IBI result shown in Fig. 1a. This IBI-fit initial guess gave nearly identical optimization results to the WCA initial guess. As we altered the pressure in the NPT optimizations, we used results from optimizations at other nearby pressures for the initial guess. We performed optimizations at $P=0.125P^*, 0.25P^*, 0.5P^*, 0.75P^*, 1.0P^*, 1.25P^*, 1.5P^*, 2.0P^*, 3.0P^*$ and $4.0P^*$. As the pressure was lowered below $P^*$, the initial guess was given by the closest optimization result at higher pressure, and as the pressure was raised above $P^*$, the initial guess corresponded to the nearest optimization result performed at lower pressure. The quality of the optimizations at $P=0.75P^*, 1.25P^*$ and $1.5P^*$ were consistent with the results presented in the main text and only omitted for clarity in the figures. 

The optimized parameters at each pressure and the density at which the validation simulations were performed are given in the table below.
\begin{table}[ht]
\label{tbl:re_parameters}
\centering
\caption{Optimal parameters for the potential form defined by Eqs.~\ref{eqn:wca}--~\ref{eqn:aux} and the corresponding density as a function of pressure.}
\begin{tabular}{  >{\centering\arraybackslash}m{1.5cm}  |  >{\centering\arraybackslash}m{1.25cm}  >{\centering\arraybackslash}m{1.25cm}  >{\centering\arraybackslash}m{1.25cm}  >{\centering\arraybackslash}m{1.25cm} >{\centering\arraybackslash}m{1.25cm}} 
\hline
Pressure & \(\beta \epsilon_{1}\) & \(\alpha_{1}\) & \(\beta \epsilon_{2}\) & \(\alpha_{2}\) & $\rho \sigma^3$ \\
\hline
$0.125P^*$ & \(1.345\) & \(1.201\) & \(0.095\) & \(2.111\) & $0.128$ \\
$0.25P^*$ & \(1.189\) & \(1.293\) & \(0.098\) & \(2.260\) & $0.144$ \\
$0.5P^*$ & \(0.980\) & \(1.441\) & \(0.121\) & \(2.461\) & $0.144$ \\
$0.75P^*$ & \(0.844\) & \(1.550\) & \(0.138\) & \(2.591\) & $0.145$ \\
$1.0P^*$ & \(0.727\) & \(1.621\) & \(0.159\) & \(2.671\) & $0.144$ \\
$1.25P^*$ & \(0.646\) & \(1.659\) & \(0.177\) & \(2.745\) & $0.144$ \\
$1.5P^*$ & \(0.567\) & \(1.682\) & \(0.194\) & \(2.799\) & $0.145$ \\
$2.0P^*$ & \(0.427\) & \(1.728\) & \(0.223\) & \(2.904\) & $0.145$ \\
$3.0P^*$ & \(0.179\) & \(1.742\) & \(0.294\) & \(3.035\) & $0.145$ \\
$4.0P^*$ & \(-0.009\) & \(1.478\) & \(0.373\) & \(3.352\) & $0.145$ \\
\hline
\end{tabular}
\end{table}


%

\end{document}